\documentclass[%
 reprint,
 amsmath,amssymb,
 aps,
 pra
]{revtex4-2}

\usepackage{graphicx}
\usepackage{dcolumn}
\usepackage{bm}

\usepackage{braket}

\begin{document}


\title{Entanglement-enhanced Synchronous differential comparison}
\author{Deshui Yu$^{1}$}
\author{Jingbiao Chen$^{2}$}
\thanks{jbchen@pku.edu.cn}
\author{Shougang Zhang$^{1,3}$}
\thanks{szhang@ntsc.ac.cn}
\affiliation{%
$^{1}$ National Time Service Center, Chinese Academy of Sciences, Xi’an 710600, China \\
$^{2}$ State Key Laboratory of Advanced Optical Communication Systems and Networks, Institute of Quantum Electronics, School of Electronics, Peking University, Beijing 100871, China\\
$^{3}$ University of Chinese Academy of Sciences, Beijing 100049, China
}%

\date{\today}

\begin{abstract}
The quantum entanglement enables the precision measurement and frequency metrology beyond the standard quantum limit that is imposed by the quantum projection noise and photon shot noise. Here we propose employing the entangled atoms in the synchronous differential measurement to enhance the sensitivity of the spatial-shift detection. Two ways of engineering the entangled atoms are studied. The synchronous comparison between two pixels within an entangled atomic cloud leads to a sensitivity enhancement factor of 1.4 over the standard quantum limit. Increasing the atom number hardly further improves the sensitivity. In contrast, the synchronous comparison between two independent pixels that are individually composed of entangled atoms allows for a strong sensitivity enhancement by a factor of, for example, 9.7 with $10^{3}$ entangled atoms in each pixel, corresponding to a reduction of the averaging time by a factor of about $10^{2}$. A large atom number may further elevate the sensitivity. Our work paves the way towards the entanglement-enhanced detection of the gravitational redshift by means of the \emph{in situ} imaging spectroscopy.
\end{abstract}

\maketitle

\section{Introduction}

Optical atomic clocks have achieved the fractional frequency instabilities and systematic uncertainties at the $10^{-18}$ level~\cite{PRL:Chou2010,Science:Hinkley2013,Nature:Bloom2014,NatPhoton:Ushijima2015,PRL:Huntemann2016,Metrologia:Bothwell2019}, towards a redefinition of the SI second based upon the optical transitions in atoms. The frequency comparison between two clocks or in a clock network lies at the heart of various applications in precision measurement and sensing on, for instance, the variation of fundamental constants such as the proton-to-electron mass ratio and the fine structure constant~\cite{PRL:Huntemann2014,PRL:Godun2014}, the gravity potential difference between remote locations (i.e., relativistic geodesy~\cite{NatPhys:Komor2014,NatPhoton:Takano2016,NatCommun:Lisdat2016}), and the interactions between atoms~\cite{Science:Swallows2011,PRL:Hutson2019,Science:Aeppli2022}. Recently, the robust comparison between two transportable optical lattice clocks has resulted in the most precise ground-based measurement of the gravitational redshift~\cite{NatPhoton:Takamoto2020}.

Thus far, the frequency instabilities of most optical clocks are still limited by the phase fluctuations in the local lasers, which are used to periodically interrogate the engineered atoms, through the Dick effect~\cite{IEEE:Santarelli1998}. A great deal of effort has been paid to suppress the local oscillator noise. It has been demonstrated that placing the high-finesse optical resonators, to which the local oscillators are prestabilized, in the cryogenic environment can vastly reduce their statistical Brownian thermal noise~\cite{NatPhoton:Kessler2012,Optica:Robinson2019}. The resultant linewidths of the prestabilized lasers may reach the millihertz level with a coherence time of up to 50 s and a fractional frequency instability of $4\times10^{-17}$ at the averaging time of 1 s~\cite{PRL:Matei2017}. Nevertheless, the complex laser-prestabilization measures add the significant complexity and volume to the clock setup.

\begin{figure}[b]
\centering
\includegraphics[width=8.5cm]{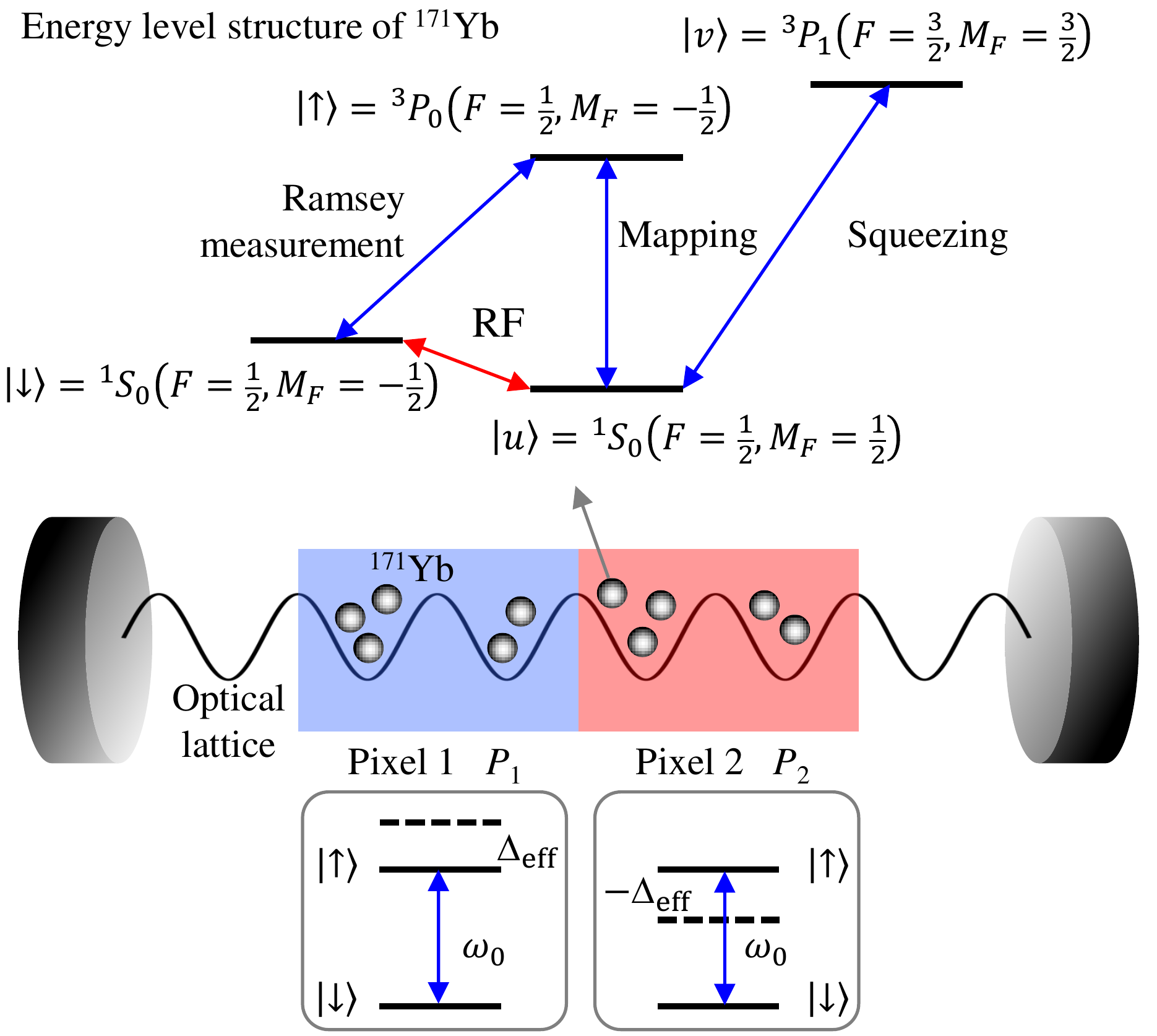}
\caption{Schematic of synchronous differential comparison. The lattice-trapped atomic cloud is divided into two pixels with the same atom number $N$. Each atom is composed of two spin states, $\ket{\downarrow}=(6s^{2})~^{1}S_{0}(F=\frac{1}{2},M_{F}=-\frac{1}{2})$ and $\ket{\uparrow}=(6s6p)~^{3}P_{0}(F=\frac{1}{2},M_{F}=-\frac{1}{2})$. The auxiliary $\ket{u}=(6s^{2})~^{1}S_{0}(F=\frac{1}{2},M_{F}=\frac{1}{2})$ and $\ket{v}=(6s6p)~^{3}P_{1}(F=\frac{3}{2},M_{F}=\frac{3}{2})$ states are used to create the entangled spins. The spatial shifts of two pixels are $\Delta_{\textrm{eff}}$ and $-\Delta_{\textrm{eff}}$, respectively.}
\label{fig1}
\end{figure}

In many practical applications, the absolute frequency measurement of the optical clocks is unnecessary. Employing the synchronous differential comparison enables the cancellation of the local oscillator noise~\cite{NatPhoton:Takamoto2011,NatPhoton:Schioppo2017} and reaching the standard quantum limit that is imposed by the quantum projection noise~\cite{NatPhoton:Oelker2020}. In particular, the synchronous clock comparison benefits from the long atom--atom coherence time (in principle, up to $10^{2}$) and has demonstrated a fractional frequency uncertainty at the $10^{-21}$ level after about $10^{2}$ hours of averaging~\cite{Nature:Zheng2022}. Such an unprecedented precision can resolve the gravitational redshift between two clocks spatially separated at the submillimetre scale~\cite{Nature:Bothwell2022}, offering the new opportunities for testing the fundamental physics. However, the long averaging time prevents the synchronous comparison from detecting relatively fast physical processes.

The standard quantum limit of the clock frequency instability may be overcome by using the entangled (correlated) atoms~\cite{NatCommun:Schulte2020,NatPhys:Schine2022,Nature:Nichol2022}. An optical lattice clock with hundreds of entangled atoms can shorten the averaging time by approximately three times, compared to the one with independent atoms~\cite{Nature:Pedrozo-Penafiel2020}. Combining the local-oscillator-noise cancellation in the synchronous comparison and the entanglement-enhanced sensitivity beyond the standard quantum limit potentially advances the timekeeping precision and stimulates various scientific applications such as the tests of the fundamental laws of physics~\cite{Science:Wciso2018} and the gravitational-wave detection~\cite{PRD:Kolkowitz2016}.

In this work, we theoretically explore the application of the entangled atoms in the synchronous differential comparison. We focus on the \emph{in situ} imaging spectroscopy of a lattice-trapped atomic cloud that is divided into two pixels. For each pixel containing $10^{3}$ atoms, the entanglement of atoms enhances the spatial-shift sensitivity by a factor of 9.7 and significantly shortens the averaging time of the synchronous measurement. Increasing the atom number further raises the metrological gain. The recent atom-optical technologies allow for testing the predictions obtained in this study.

\section{Physical system}

Figure~\ref{fig1} illustrates the schematic diagram of the physical system. An ensemble of $^{171}$Yb atoms are tightly confined in a one-dimensional magic-wavelength (759 nm) optical lattice~\cite{SciRep:Gao2018}. The atomic cloud is divided into two groups, corresponding to two pixels in the \emph{in situ} imaging spectroscopy~\cite{PRL:Marti2018}. We assume that two pixels contain the same number $N$ of atoms. Each atom is modelled as a spin composed of the clock $\ket{\downarrow}=(6s^{2})~^{1}S_{0}(F=\frac{1}{2},M_{F}=-\frac{1}{2})$ and $\ket{\uparrow}=(6s6p)~^{3}P_{0}(F=\frac{1}{2},M_{F}=-\frac{1}{2})$ states with the transition frequency of $\omega_{0}=2\pi\times519$ THz. Other atomic levels and transitions shown in Fig.~\ref{fig1} are used for creating the entangled atoms (see below). We neglect the spontaneous decay of the spins since the time scale of interest in this work is much shorter than the natural lifetime (over 16 s~\cite{PRA:Porsev2004}) of the $\ket{\uparrow}$ state.

The sub-Hilbert space for the $k$-th pixel ($k=1,2$) is spanned by the collective spin basis $\ket{J_{k},M_{k}}$ (i.e., the exchange-symmetric Dicke manifold), where $J_{k}=\frac{N}{2}$ denotes the total spin angular momentum quantum number and $M_{k}$ corresponds to the projection quantum number along the quantization axis. The angular momentum vector operator for the $k$-th pixel is written as $\hat{\textbf{J}}_{k}=\sum_{\mu=x,y,z}\hat{J}_{\mu,k}\hat{\textbf{e}}_{\mu}$ with three components $\hat{J}_{\mu,k}=\frac{1}{2}\sum_{n=1}^{N}\hat{\sigma}_{\mu}^{(k,n)}$ and the Pauli matrices $\hat{\sigma}_{x}^{(k,n)}=(\ket{\downarrow}\bra{\uparrow}+\ket{\uparrow}\bra{\downarrow})_{k,n}$, $\hat{\sigma}_{y}^{(k,n)}=i(\ket{\downarrow}\bra{\uparrow}-\ket{\uparrow}\bra{\downarrow})_{k,n}$, and $\hat{\sigma}_{z}^{(k,n)}=(\ket{\uparrow}\bra{\uparrow}-\ket{\downarrow}\bra{\downarrow})_{k,n}$ for the $n$-th spin in the $k$-th pixel. We have $\hat{\bf{J}}_{k}^{2}\ket{J_{k},M_{k}}=J_{k}(J_{k}+1)\ket{J_{k},M_{k}}$ and $\hat{J}_{z,k}\ket{J_{k},M_{k}}=M_{k}\ket{J_{k},M_{k}}$. The Hilbert space for the entire spin system is then spanned by $\ket{J_{1},M_{1};J_{2},M_{2}}$, which can be simplified as $\ket{M_{1},M_{2}}$ due to $J_{1}=J_{2}=\frac{N}{2}$.

In the $k$-th pixel, the spin population in $\ket{\uparrow}$ is given by $P_{k}=\langle\hat{P}_{k}\rangle=\bra{\psi}\hat{P}_{k}\ket{\psi}$ with the system state $\psi$ and the projection operator $\hat{P}_{k}=\hat{J}_{z,k}+\frac{N}{2}$. The total projection operator is written as $\hat{P}=\hat{P}_{1}+\hat{P}_{2}$ and $\hat{P}_{d}=\hat{P}_{2}-\hat{P}_{1}$ corresponds to the population-difference projection operator. The standard deviation of an arbitrary operator $\hat{Q}$ is evaluated through
\begin{equation}
\Delta Q=\sqrt{\langle\hat{Q}^{2}\rangle-\langle\hat{Q}\rangle^{2}}.
\end{equation}

The spin system experiences the extra spatially dependent frequency shifts that may be caused by, for example, the residual magnetic field gradient, the lattice-induced shifts, and the gravitational redshift. We assume that the inhomogeneity of the spins within each pixel is negligible due to the small pixel size. By contrast, in order to model the relative frequency difference between two pixels, the effective spatial frequency shifts $\pm\Delta_{\textrm{eff}}$ are introduced to the pixels, respectively (see Fig.~\ref{fig1}).

\begin{figure}
\centering
\includegraphics[width=8.5cm]{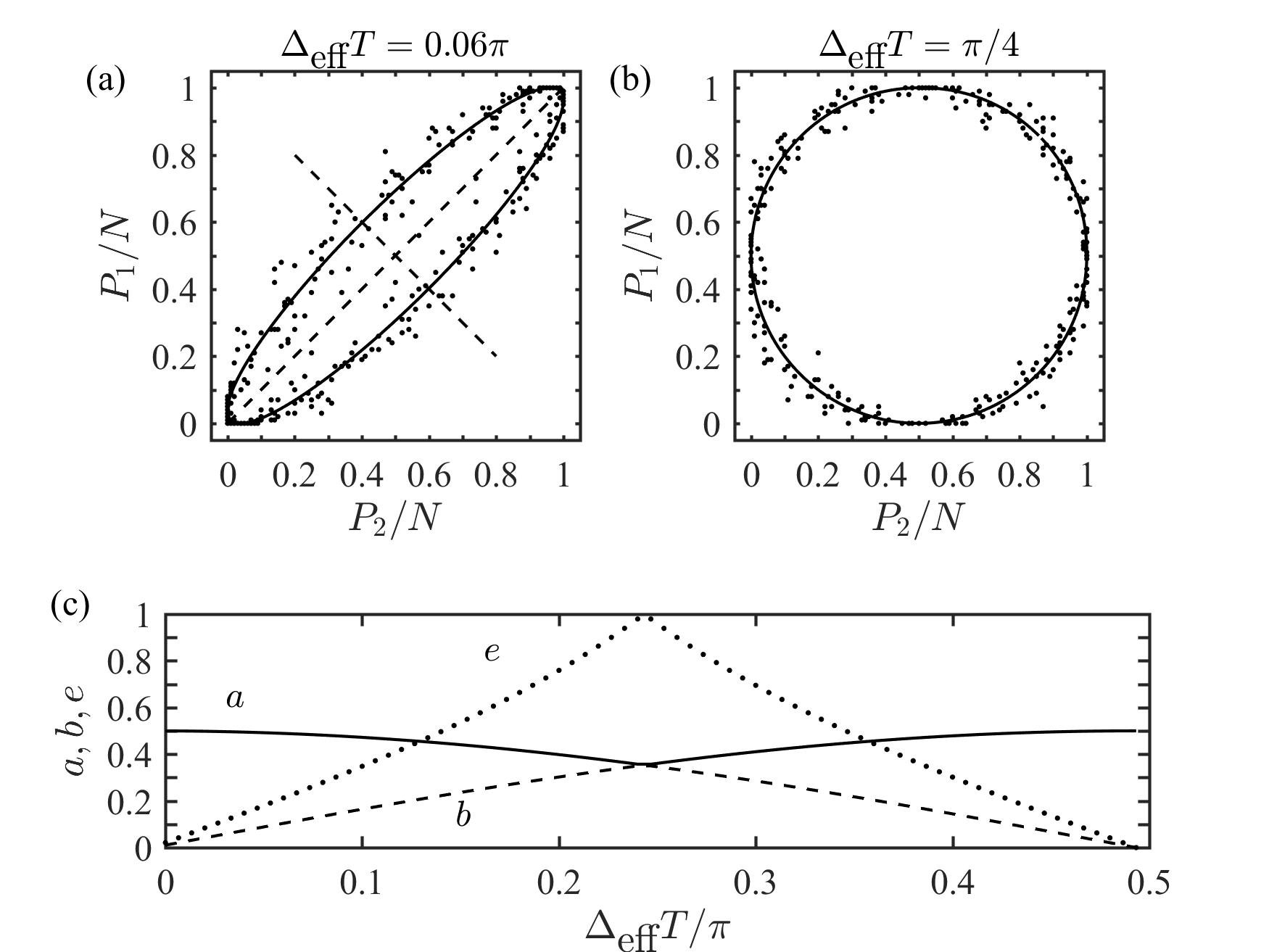}
\caption{Synchronous differential comparison with independent spins. (a) and (b) Parametric plots of the excitation fraction of two pixels for different $\Delta_{\textrm{eff}}T$. Symbols: Monte Carlo simulation. Solid lines: analytical results. (c) Dependence of the lengths of half-major $a$ (solid) and half-minor $b$ (dashed) axes and the ellipticity $e=b/a$ (dotted) on $\Delta_{\textrm{eff}}T$.}
\label{fig2}
\end{figure}

We perform the Ramsey measurement. The spin system is initially prepared in $\psi_{0}=\ket{M_{1}=-\frac{N}{2},M_{2}=-\frac{N}{2}}$, i.e., all spins are in $\ket{\downarrow}$. Two light $\frac{\pi}{2}$-pulses with an interval of $T$ are successively applied to excite the spins. The phase difference between two pulses is $\phi$. The populations of two pixels in $\ket{\uparrow}$, $P_{1}$ and $P_{2}$, are then measured and the population difference $P_{d}=P_{2}-P_{1}$ is computed. The mathematical treatment of the Ramsey measurement is listed in Appendix A.

\section{Independent spins}

For the system composed of independent spins, the dependence of the Ramsey excitation fractions of two pixels on the phase $\phi$ are expressed as
\begin{subequations}\label{ellipse_eq}
\begin{eqnarray}
\frac{P_{1}}{N}&=&\frac{1}{2}+\frac{C}{2}\cos(\phi+\Delta_{\textrm{eff}}T),\\
\frac{P_{2}}{N}&=&\frac{1}{2}+\frac{C}{2}\cos(\phi-\Delta_{\textrm{eff}}T).
\end{eqnarray}
\end{subequations}
Due to the absence of the spin decay, the Ramsey fringe contrast $C$ is equal to unity. We make the parametric plot that graphs $P_{1}$ and $P_{2}$ on a coordinate system. The plot exhibits an ellipse with the lengths of major and minor axes of $2a$ and $2b$, respectively (see Fig. \ref{fig2}a and b). Varying the spatial-shift-induced phase $\Delta_{\textrm{eff}}T$ changes the ellipticity $e=b/a$ of the ellipse (see Fig. \ref{fig2}c). The parametric plot becomes a circle, i.e., $e=1$, when $\Delta_{\textrm{eff}}T=\pi/4$. The orientation angle of the ellipse, i.e., the angle of the semi-major axis that is measured counter-clockwise from the positive horizontal axis, is equal to $\frac{\pi}{4}$ or $-\frac{\pi}{4}$, depending on if $\Delta_{\textrm{eff}}T$ is less than $\pi/4$.

We are interested in detecting the small spatial shift $\Delta_{\textrm{eff}}$. In the limit of $\Delta_{\textrm{eff}}T\sim0$, Eq. (\ref{ellipse_eq}) leads to
\begin{equation}\label{delta_eff_eq}
\Delta_{\textrm{eff}}=\frac{P_{d}}{NT\sin\phi},
\end{equation}
with $P_{d}=\langle\hat{P}_{d}\rangle$. Thus, one may evaluate $\Delta_{\textrm{eff}}$ through measuring the Ramsey excitation difference $P_{d}$ between two pixels. Usually, the phase difference between two light $\frac{\pi}{2}$-pulses is set as $\phi=\frac{\pi}{2}$. Then, the uncertainty $\sigma(\Delta_{\textrm{eff}})$ of the spatial-shift measurement is determined by the standard deviation $\Delta P_{d}$ of $P_{d}$,
\begin{eqnarray}\label{std_Pd}
\nonumber\sigma(\Delta_{\textrm{eff}})&=&\Delta P_{d}/(NT)\\
&=&(NT)^{-1}\sqrt{(\Delta P_{1})^2+(\Delta P_{2})^2-2G(P_{1},P_{2})},~~~~
\end{eqnarray}
where we have defined the correlation function 
\begin{equation}
G(P_{1},P_{2})=\langle\hat{P}_{1}\hat{P}_{2}\rangle-\langle\hat{P}_{1}\rangle\langle\hat{P}_{2}\rangle.
\end{equation}
Actually, Eq.~(\ref{std_Pd}) represents the sensitivity of the spatial-shift measurement. For independent spins, two pixels are uncorrelated and one has $G(P_{1},P_{2})=0$. We neglect all technique noise sources. The standard deviation of the Ramsey measurement is completely caused by the quantum projection noise, $\Delta P_{1}=\Delta P_{2}=\frac{\sqrt{N}}{2}$~\cite{PRA:Itano1993}, and we obtain
\begin{equation}
\sigma(\Delta_{\textrm{eff}})=(\sqrt{2N}T)^{-1}.
\end{equation}
Extending the free-evolution time $T$ suppresses the uncertainty $\sigma(\Delta_{\textrm{eff}})$. Nevertheless, the spin decay aggravates the uncertainty for a large $T$.

An arbitrary state of the ensemble of independent spins may be written as~\cite{PhysRep:Ma2011}
\begin{equation}
\psi_{\textrm{CSS}}(\theta,\textbf{n})=e^{-i\theta\textbf{n}\cdot(\hat{\textbf{J}}_{1}+\hat{\textbf{J}}_{2})}\psi_{0},
\end{equation}
i.e., rotating the initial state $\psi_{0}$ around the axis $\textbf{n}$ by an angle of $\theta$. The expectation value of the total projection operator is $P=\langle\hat{P}\rangle=N(1-\cos\theta)$ and the corresponding standard deviation is $\Delta P=\sqrt{N/2}|\sin\theta|$. The $\psi_{\textrm{CSS}}(\theta,\textbf{n})$ state is usually referred to as the coherent spin state (CSS), in which the standard deviation $\Delta P$ of the quantum projection noise scales as $\sqrt{N}$, i.e., the standard quantum limit~\cite{PRA:Santarelli1999}. For the Ramsey measurement, the spin system is in $\psi_{\textrm{CSS}}(\theta=\frac{\pi}{2},\textbf{n}=\textbf{e}_{x})$ before the second light $\frac{\pi}{2}$-pulse is launched.

\section{Entangled spin system}

\begin{figure}
\centering
\includegraphics[width=8.5cm]{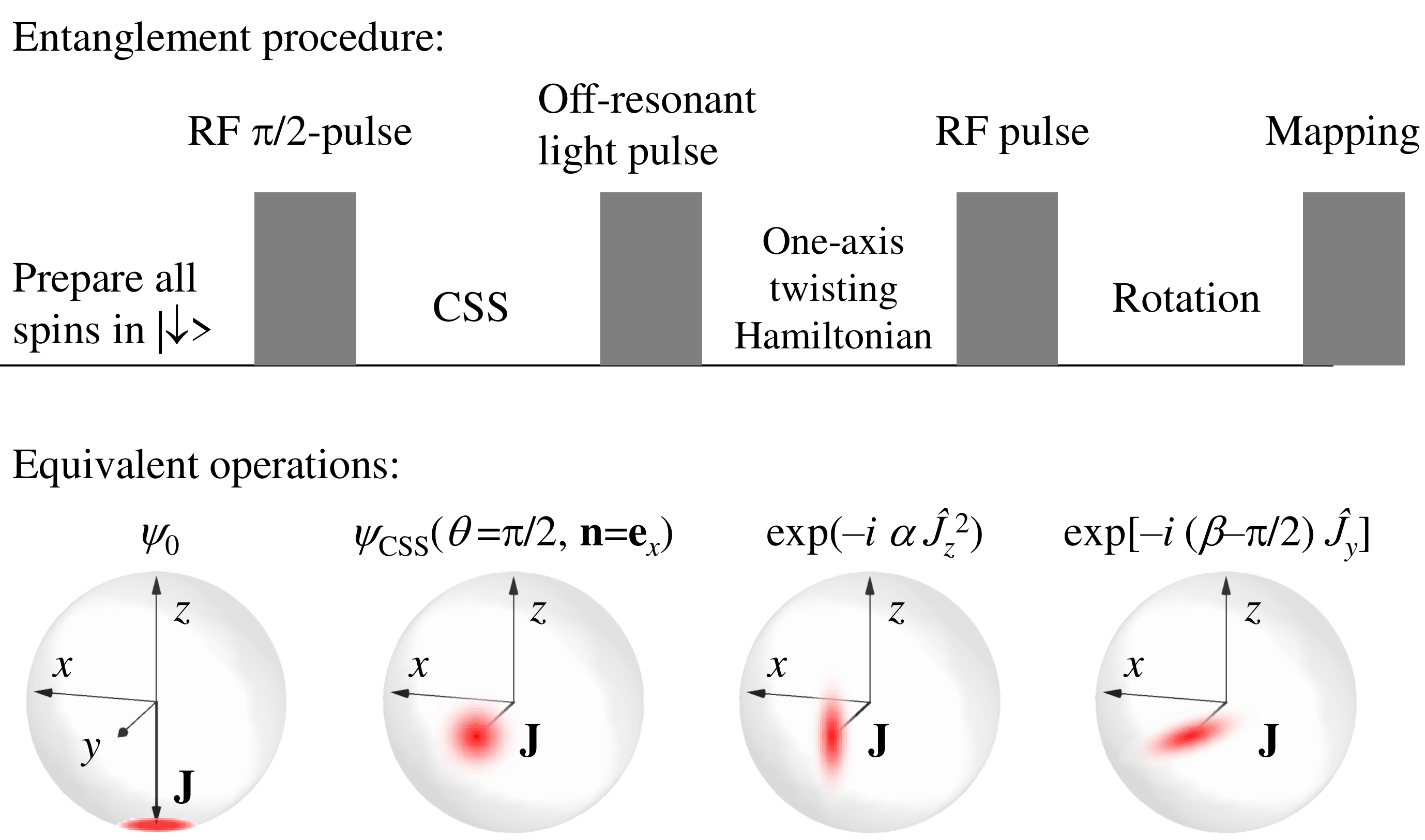}
\caption{Procedure of creating the entangled spins. Upper panel: The sequence of RF and light pulses driving the $\ket{\downarrow}-\ket{u}$, $\ket{u}-\ket{v}$, and $\ket{u}-\ket{\uparrow}$ transitions in Fig.~\ref{fig1}. Lower panel: The equivalent operations performed on the spin system.}
\label{fig3}
\end{figure}

It has been pointed out that the standard quantum limit may be overcome by employing the entangled (correlated) spins~\cite{PRA:Wineland1992,PRA:Wineland1994}. Various methods of generating the entanglement between atoms have been proposed~\cite{PRA:Kitagawa1993,PRA:SchleierSmith2010,PRL:Perlin2020} and demonstrated~\cite{Nature:Pedrozo-Penafiel2020,Science:Lange2018,Nature:Bao2020}. Here, we follow the approach based on the one-axis twisting Hamiltonian~\cite{PRA:SchleierSmith2010,Nature:Pedrozo-Penafiel2020}. Equations~(\ref{delta_eff_eq}) and (\ref{std_Pd}) are still valid for the entangled spins. As we will see below, the resultant nonzero inter-pixel correlation $G({P_{1},P_{2}})$ may suppress the deviation $\Delta P_{d}$ of the Ramsey excitation difference below the standard quantum limit.

To create the entangled atoms, two auxiliary states, $\ket{u}=(6s^{2})~^{1}S_{0}(F=\frac{1}{2},M_{F}=\frac{1}{2})$ and $\ket{v}=(6s6p)~^{3}P_{1}(F=\frac{3}{2},M_{F}=\frac{3}{2})$, are involved (see Fig.~\ref{fig1}), where $\ket{u}$ acts as a role similar to $\ket{\uparrow}$. The specific procedure can be summarized as follows (see Fig.~\ref{fig3}): All spins are initialized in $\ket{\downarrow}$. A radiofrequency (RF) $\frac{\pi}{2}$-pulse is used to create a CSS between $\ket{\downarrow}$ and $\ket{u}$. This step corresponds to the preparation of the spin system in $\psi_{\textrm{CSS}}(\theta=\frac{\pi}{2},\textbf{n}=\textbf{e}_{x})$. Subsequently, an off-resonant light pulse drives the $\ket{u}-\ket{v}$ transition through the optical cavity, introducing the cavity-mediated interactions between atoms. This step corresponds to the evolution of the spin system under the one-axis twisting Hamiltonian $\hat{J}_{z}^{2}$ for a duration $\alpha$. Usually, the spin-echo technique is used to eliminate the acquired linear phase shift~\cite{Nature:Pedrozo-Penafiel2020}. Then, another RF pulse is applied to rotate the system composed of $\ket{\downarrow}$ and $\ket{u}$ around the $y$-axis by an angle $(\beta-\frac{\pi}{2})$, equivalent to imposing the same operation on the spin system. Finally, a light $\pi$-pulse is launched to map the population in $\ket{u}$ to $\ket{\uparrow}$. In theory, the whole procedure is described as
\begin{eqnarray}\label{spin_squeezed_state_1}
\nonumber\psi_{\textrm{SSS1}}(\alpha,\beta)&=&e^{-i(\beta-\pi/2)(\hat{J}_{y,1}+\hat{J}_{y,2})}e^{-i\alpha(\hat{J}_{z,1}+\hat{J}_{z,2})^{2}}\\
&&\times\psi_{\textrm{CSS}}(\theta=\pi/2,\textbf{n}=\textbf{e}_{x}).
\end{eqnarray}
After using the second light $\frac{\pi}{2}$-pulse, one may evaluate the Ramsey excitation refractions of two pixels. It is worth noting that unlike the application of the entangled spins in quantum metrology (see Appendix B), the spin entanglement here is applied to suppress the standard deviation $\Delta P_{d}$ of the excitation difference $P_{d}$ between two pixels so as to reduce the measurement uncertainty $\sigma(\Delta_{\textrm{eff}})$ of the spatial shift $\Delta_{\textrm{eff}}$. Equation~(\ref{spin_squeezed_state_1}) is referred to as the squeezed spin state (SSS).

\begin{figure}
\centering
\includegraphics[width=9cm]{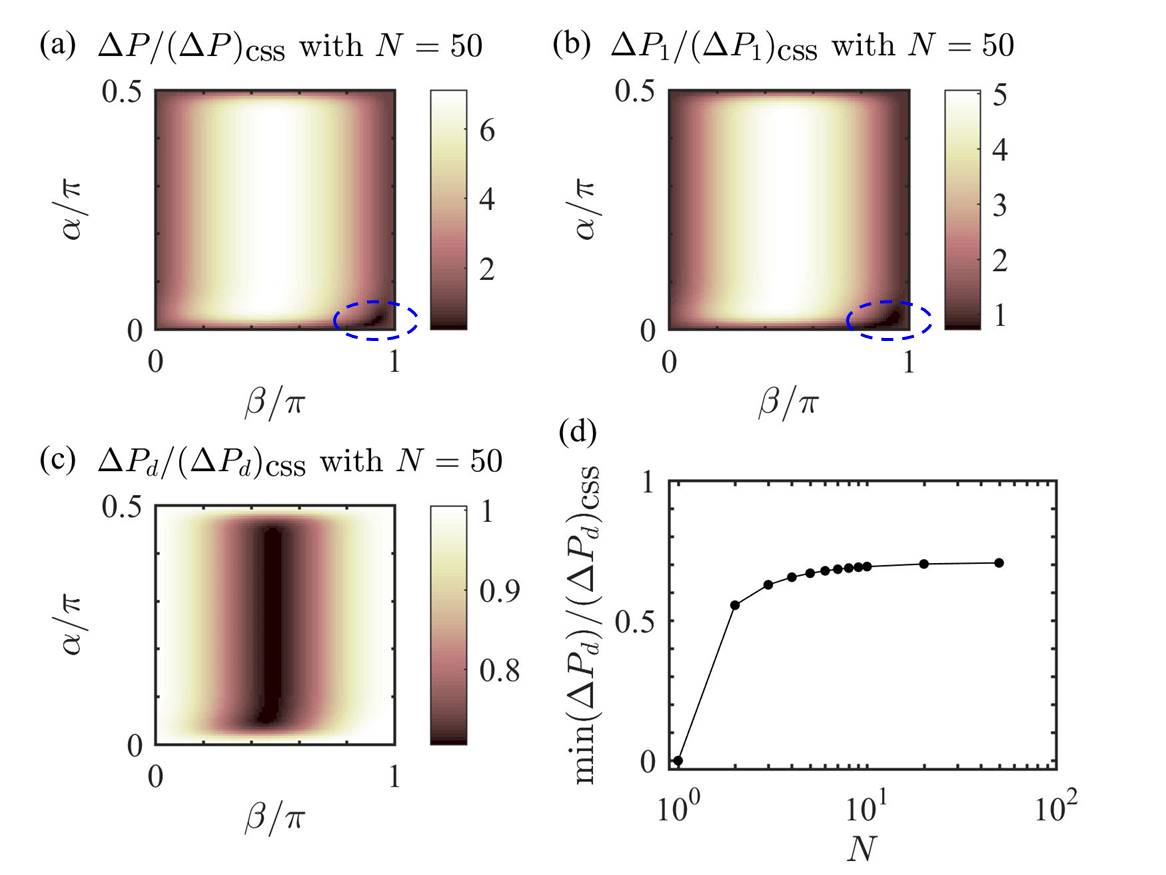}
\caption{Standard deviations $\Delta P$ (a), $\Delta P_{1}=\Delta P_{2}$ (b), and $\Delta P_{d}$ (c) of the entangled spins as a function of the angles $\alpha$ and $\beta$ with the atom number $N=50$ of each pixel. The dashed circles in (a) and (b) show the small regimes where $\Delta P<(\Delta P)_{\textrm{CSS}}$ and $\Delta P_{k=1,2}<(\Delta P_{k})_{\textrm{CSS}}$. (d) Dependence of the minimum of $\Delta P_{d}$ on $N$.}
\label{fig4}
\end{figure}

Figure \ref{fig3}a plots the deviation $\Delta P$ of the total projection operation $\hat{P}$ (i.e., the Ramsey excitation of the whole spin system) as a function of the angles $\alpha$ and $\beta$. It is found that within the most $(\alpha,\beta)$ regime, $\Delta P$ exceeds the corresponding standard quantum limit $(\Delta P)_{\textrm{CSS}}=\sqrt{N/2}$. That is, the entanglement does not always suppress the projection noise. Nevertheless, there is the certain $(\alpha,\beta)$ regime, within which one has $\Delta P<(\Delta P)_{\textrm{CSS}}$. Since the individual pixels are parts of the whole entangled spin system, the dependence of $\Delta P_{k=1,2}$ of $\hat{P}_{k}$ on $(\alpha,\beta)$ is similar to that of $\Delta P$ (see Fig. \ref{fig3}b).

Our goal is to reduce the deviation $\Delta P_{d}$ of the excitation difference $P_{d}$ [see Eq.~(\ref{std_Pd})], rather than the suppression of $\Delta P$ or $\Delta P_{k=1,2}$. Interestingly, as shown in Fig. \ref{fig4}c, $\Delta P_{d}$ is always less than or equal to the standard quantum limit $(\Delta P_{d})_{\textrm{CSS}}=\sqrt{N/2}$. The minimum of $\Delta P_{d}$ reaches zero, $\textrm{min}(\Delta P_{d})=0$, when each pixel contains only one spin, i.e., $N=1$. Actually, this situation is analogous to the recent experiment in~\cite{Nature:Nichol2022}, where two single ion clocks are fully entangled, i.e., $\psi_{\textrm{SSS1}}=\frac{1}{\sqrt{2}}(\ket{M_{1}=-\frac{1}{2},M_{2}=-\frac{1}{2}}+\ket{M_{1}=\frac{1}{2},M_{2}=\frac{1}{2}})$. As $N$ is increased, the ratio $\textrm{min}(\Delta P_{d})/(\Delta P_{d})_{\textrm{CSS}}$ approaches a saturation value of $\sim0.7$ (see Fig. \ref{fig4}d), corresponding to a metrological gain of 1.5 decibels and a reduction of the averaging time by a factor of 2.

Indeed, the suppression of $\Delta P_{d}$ is attributed to the fact $G(P_{1},P_{2})\geq0$ [see Eq. (\ref{std_Pd})]. Comparing Fig. \ref{fig4}b and c, one finds that the strong suppression of $\Delta P_{d}$ occurs within the $(\alpha,\beta)$ regime where $\Delta P_{k=1,2}$ are strongly enhanced. Thus, despite the strong quantum fluctuations in the Ramsey measurement of individual pixels, the large inter-pixel correlation (i.e., the entanglement between two pixels) still ensures the suppression of the quantum noise in the synchronous differential measurement.

\begin{figure}[b]
\centering
\includegraphics[width=9cm]{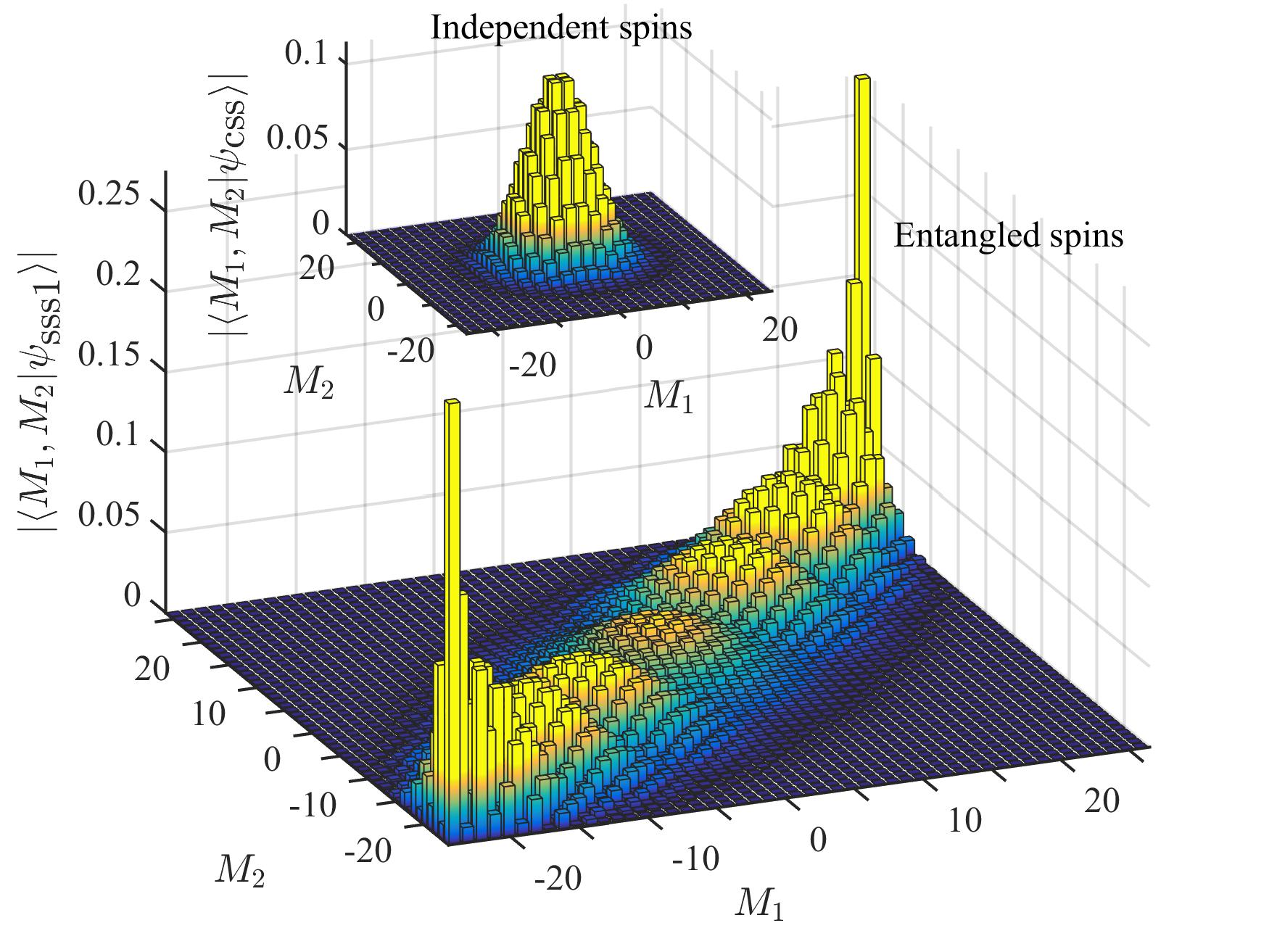}
\caption{Population of the entangled $\psi_{\textrm{SSS1}}(\alpha,\beta)$ state in different $\ket{M_{1},M_{2}}$ states. Here, $\psi_{\textrm{SSS1}}(\alpha,\beta)$ is the state in which $\Delta P_{d}/(\Delta P_{d})_{\textrm{CSS}}$ is minimized. The atom number of each pixel is $N=50$. Inset: Population distribution for the independent spins, i.e., the spin system is in $\psi_{\textrm{CSS}}(\theta=\frac{\pi}{2},\textbf{n}=\textbf{e}_{x})$.}
\label{fig5}
\end{figure}

In order to gain insight into the entanglement between two pixels, we compute the distribution of the entangled $\psi_{\textrm{SSS1}}$ state in different $\ket{M_{1},M_{2}}$ states. Here, $\psi_{\textrm{SSS1}}$ corresponds to the state in which $\Delta P_{d}/(\Delta P_{d})_{\textrm{CSS}}$ reaches its minimum. As depicted in Fig.~\ref{fig5}, the spin system is mainly populated around the diagonal line with $M_{1}=M_{2}$ in the $(M_{1},M_{2})$ regime, denoting the synchronized excitation of two pixels. In addition, the distribution is maximized at the $\ket{M_{1}=\frac{N}{2},M_{2}=\frac{N}{2}}$ and $\ket{M_{1}=-\frac{N}{2},M_{2}=-\frac{N}{2}}$ states, characterizing the cooperative behavior of spins. According to the distribution shown in Fig.~\ref{fig5}, one may numerically simulate the Ramsey excitation fractions $P_{1,2}$ of two pixels by using the Monte Carlo method. In contrast, the system composed of independent spins is mainly populated around $\ket{M_{1}=0,M_{2}=0}$ and the distribution presents an isotropic Gaussian pattern (see the inset in Fig.~\ref{fig5}), indicating the single spin behavior.

\section{Two independent pixels with each individual composed of entangled spins}

\begin{figure}
\centering
\includegraphics[width=9cm]{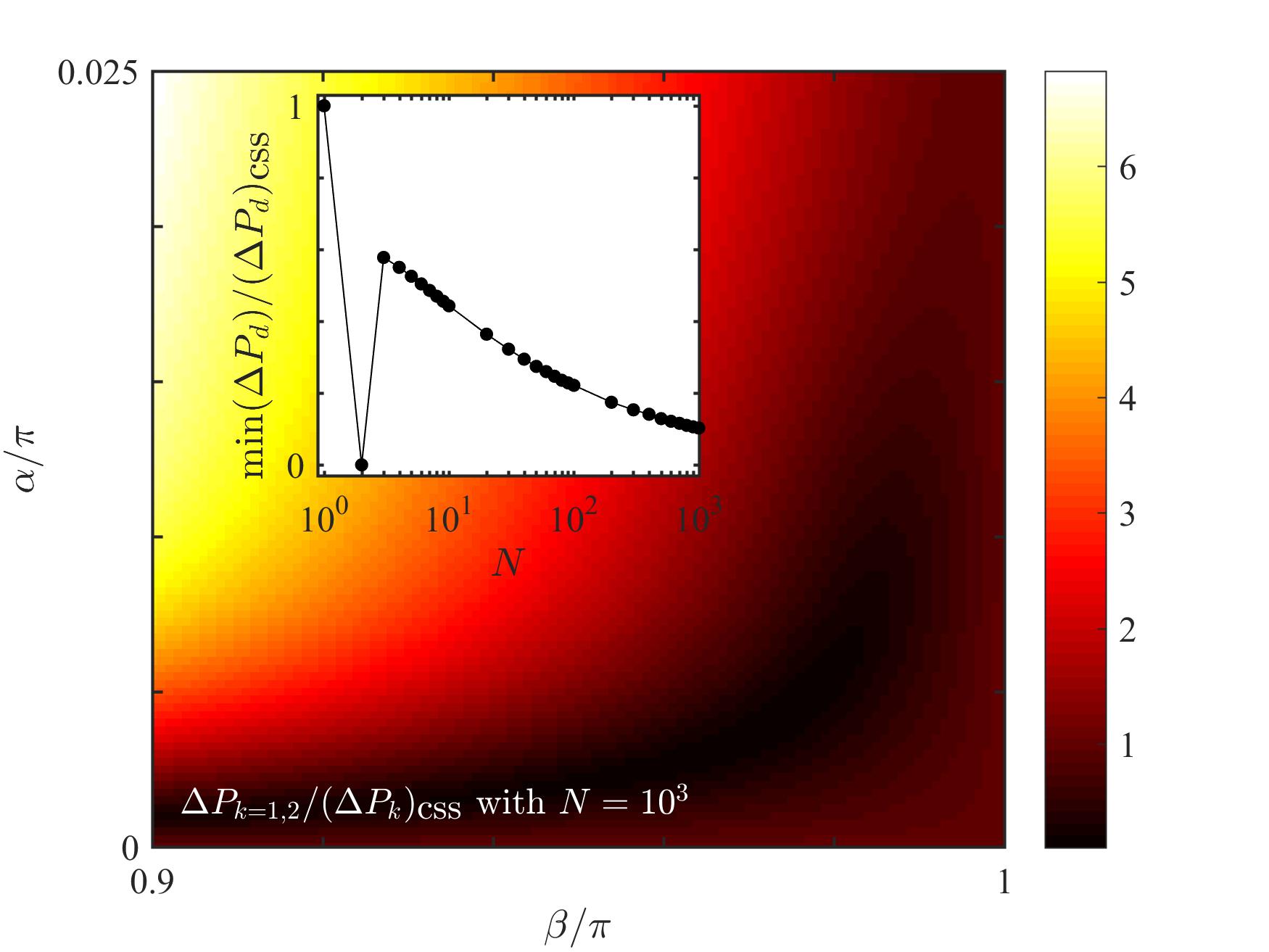}
\caption{Dependence of the deviation $\Delta P_{k=1,2}$ on the angles $\alpha$ and $\beta$, where the system is in $\psi_{\textrm{SSS2}}(\alpha,\beta)$ before the second light $\frac{\pi}{2}$-pulse is launched in the Ramsey measurement. The number of the spins in each pixel is $N=10^{3}$. Inset: the ratio of the minimum of $\Delta P_{d}$ to $(\Delta P_{d})_{\textrm{CSS}}$ changing with $N$.}
\label{fig6}
\end{figure}

In last section, we have discussed the synchronous comparison with the entanglement of all spins, where the positive inter-pixel correlation reduces the deviation $\Delta P_{d}$ of the Ramsey excitation difference. From Eq.~(\ref{std_Pd}), it is seen that $\Delta P_{d}$ can be also suppressed by reducing $\Delta P_{1,2}$ even when the inter-pixel correlation $G(P_{1},P_{2})$ vanishes, i.e., $\Delta P_{d}=\sqrt{2}\Delta P_{1}<(\Delta P_{d})_{\textrm{CSS}}$. Here, we have used $\Delta P_{1}=\Delta P_{2}$. Motivated by this, we further consider the following situation, where two pixels are independent while the spins in each pixel are entangled through the approach shown in Fig.~\ref{fig3}. Thus, the system is in
\begin{eqnarray}
\nonumber\psi_{\textrm{SSS2}}(\alpha,\beta)&=&\prod_{k=1,2}e^{-i(\beta-\pi/2)\hat{J}_{y,k}}e^{-i\alpha\hat{J}_{z,k}^{2}}\\
&&~~~~~\times\psi_{\textrm{CSS}}(\theta=\pi/2,\textbf{n}=\textbf{e}_{x}),
\end{eqnarray} 
before the second light $\frac{\pi}{2}$-pulse is launched in the Ramsey measurement. Such an entanglement scheme may be implemented by using the optical tweezer clock technique~\cite{PRX:Madjarov2019,Science:Norcia2019,PRL:Covey2019}.

Figure~\ref{fig6} displays the dependence of $\Delta P_{k=1,2}$ on the angles $\alpha$ and $\beta$. As is expected, within the certain $(\alpha,\beta)$ regime the entanglement of spins in each pixel may reduce $\Delta P_{k=1,2}$ below that of the pixels composed of independent spins. The inset in Fig.~\ref{fig6} plots the minimum of the resultant deviation $\Delta P_{d}$ of the Ramsey excitation difference as a function of the number $N$ of the spins in each pixel. In the simplest case with $N=1$, one has $\Delta P_{d}=(\Delta P_{d})_{\textrm{CSS}}$. For $N=2$, the squeezed spin state $\psi_{\textrm{SSS2}}$ takes the form of $\ket{\Phi}_{1}\otimes\ket{\Phi}_{2}$ with the Bell state $\ket{\Phi}_{k}=\frac{1}{\sqrt{2}}(\ket{\downarrow\downarrow}-\ket{\uparrow\uparrow})_{k}$ of the $k$-th pixel and the minimum of $\Delta P_{d}$ reaches zero. Recently, the enhanced metrological stability has been demonstrated by using the long-lived Bell states~\cite{NatPhys:Schine2022}. Unlike the system in $\psi_{\textrm{SSS1}}$, increasing the spin number $N$ may further suppress the ratio $\textrm{min}(\Delta P_{d})/(\Delta P_{d})_{\textrm{CSS}}$ for the system in $\psi_{\textrm{SSS2}}$ (see Fig.~\ref{fig4}d and the inset in Fig.~\ref{fig6}). For example, $\psi_{\textrm{SSS2}}$ leads to $\textrm{min}(\Delta P_{d})/(\Delta P_{d})_{\textrm{CSS}}\approx0.27$ (0.1) when $N=50$ ($10^{3}$), corresponding to a spatial-shift sensitivity enhancement of 5.6 (9.9) decibels. 

\section{Allan deviation}

\begin{figure}[b]
\centering
\includegraphics[width=9cm]{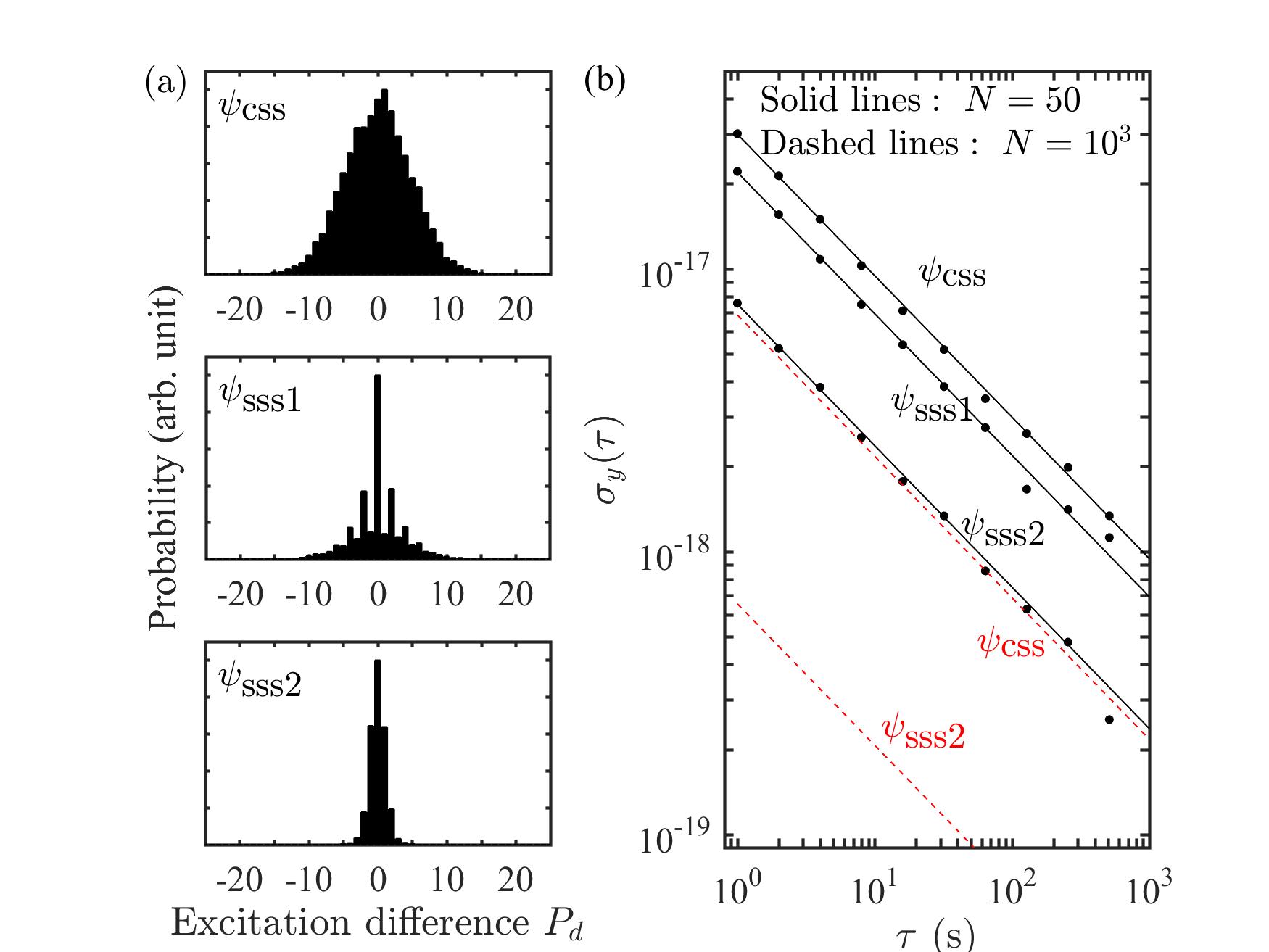}
\caption{Monte Carlo simulation. (a) Distributions of the Ramsey excitation difference $P_{d}$ between two pixels for the spin system in different states. For all plots, the spin number is $N=50$. (b) Allan deviation $\sigma_{y}(\tau)$ of the synchronous differential measurement of the spatial shift $\Delta_{\textrm{eff}}$. Symbols: numerical results with $N=50$. Solid lines: curve fitting. Dashed line: analytical results with $N=10^{3}$.}
\label{fig7}
\end{figure}

Finally, we consider the stability (i.e., Allan deviation) of the synchronous differential measurement of the spatial shift $\Delta_{\textrm{eff}}$. We numerically simulate the pixel populations $P_{1,2}$ by using the Monte Carlo method (see Appendix~\ref{appendix_ramsey_measurement}) and compute the Ramsey excitation difference $P_{d}$. Figure~\ref{fig7}a shows the distributions of $P_{d}$ for the spin system in $\psi_{\textrm{CSS}}$, $\psi_{\textrm{SSS1}}$, and $\psi_{\textrm{SSS2}}$, respectively. It is seen that the $\psi_{\textrm{CSS}}$ state leads to a Gaussian distribution of $P_{d}$ with the deviation of $\Delta P_{d}=\sqrt{N}/2$. In contrast, $P_{d}$ for the system in $\psi_{\textrm{SSS1}}$ has a distribution width narrower than that of $\psi_{\textrm{CSS}}$. The distribution width is further narrowed when the system is in $\psi_{\textrm{SSS2}}$.

According to Eq.~(\ref{delta_eff_eq}), one may compute the spatial shift $\Delta_{\textrm{eff}}$ from the measurement of $P_{d}$. This $\Delta_{\textrm{eff}}$ corresponds to the mean value of the spatial shift averaged over a cycle time $T_{c}$. Each cycle is composed of the preparation of the lattice-trapped spins, the generation of the entanglement between spins, and the Ramsey measurement. Typically, the time duration for creating the entangled spins is of the order of 0.1 s \cite{Nature:Pedrozo-Penafiel2020} and the Ramsey time $T$ is of the order of 1 s. Thus, $T_{c}$ approximates $T$. Performing the simulation repeatedly, one obtains a frequency series $\{\Delta_{\textrm{eff},k},k\in Z\}$, where the index $k$ denotes the $k$-th measurement cycle. The Allan deviation is then given by
\begin{equation}
\sigma_{y}(\tau=nT_{c})=\omega_{0}^{-1}\sqrt{\langle(\bar{\Delta}_{\textrm{eff},m+1}-\bar{\Delta}_{\textrm{eff},m})^{2}\rangle},
\end{equation}
with $\bar{\Delta}_{\textrm{eff},m}=\frac{1}{n}\sum_{k=(m-1)n+1}^{mn}\Delta_{\textrm{eff},k}$ averaged over the period and $n\in Z$.

The numerical results with $T_{c}=1$ s are shown in Fig.~\ref{fig7}b. The entanglement of spins improves the stability of the measurement of $\Delta_{\textrm{eff}}$. For $N=50$, the Allan deviation for the system in $\psi_{\textrm{CSS}}$ is evaluated to be $\sigma_{y}(\tau)=3.0\times10^{-17}/\sqrt{\tau}$. Preparing the system in $\psi_{\textrm{SSS1}}$ leads to $\sigma_{y}(\tau)=2.2\times10^{-17}/\sqrt{\tau}$. In contrast, the Allan deviation for the system in $\psi_{\textrm{SSS2}}$ is $\sigma_{y}(\tau)=7.5\times10^{-18}/\sqrt{\tau}$, denoting a metrological gain of 5.6 decibels over the standard quantum limit and a reduction of the averaging time by a factor greater than 14. The analytic expression of the Allan deviation takes the form
\begin{eqnarray}
\nonumber\sigma_{y}(\tau)&=&\frac{\sigma(\Delta_{\textrm{eff}})}{\omega_{0}}\sqrt{\frac{T_{c}}{\tau}}\\
&\approx&\frac{\Delta P_{d}/N}{2\omega_{0}\sqrt{T_{c}}}\frac{1}{\sqrt{\tau}}.
\end{eqnarray}
Since $\Delta P_{d}$ for the system in $\psi_{\textrm{SSS2}}$ decreases with the atom number $N$ (see the inset in Fig.~\ref{fig6}), one may increase $N$ to enhance the measurement stability. When $N=10^{3}$, we obtain $\sigma_{y}(\tau)=7.0\times10^{-19}/\sqrt{\tau}$, one order of magnitude better than~\cite{Nature:Bothwell2022}, by employing $\psi_{\textrm{SSS2}}$ (see Fig.~\ref{fig7}b). For two pixels with an effective separation of 10 $\mu$m, the gravitational redshift at the Earth’s surface causes a fractional frequency difference of $1.09\times10^{-21}$, which can be resolved when $\tau=100$ h. In addition, extending the free-evolution time $T$ also improves the stability $\sigma_{y}(\tau)$. However, for a large $T$, the spontaneous decay of $\ket{\uparrow}$ degrades the contrast of the Ramsey measurement and the relevant effects should be taken into account.

\section{Conclusion}

In summary, we have investigated the application of the entanglement of atoms in the synchronous differential comparison. The resultant suppression of the deviation of the Ramsey excitation difference between two pixels leads to an enhanced sensitivity of the spatial-shift detection and a strong reduction of the averaging time. The proposed metrology scheme is feasible by means of the recent atom-optical techniques and will advance the tests of fundamental physics.

In this study, we have assumed two pixels have the same atom number. However, in practice this cannot be ensured when preparing the sample. In addition, the total number of atoms veries for different samples. According to the experiment in~\cite{Nature:Pedrozo-Penafiel2020}, the standard deviation $\sigma_{N}$ of the atom number whose mean value is $N=350$ can be controlled to $\sigma_{N}=40$. In contrast, the expeiment in~\cite{NatPhoton:Takamoto2011} has $N\approx10^{3}$ with $\sigma_{N}$ of the order of 10. Assuming a deviation of $\sigma_{N}/N=0.01$, we estimate that the fluctuations of the atom number cause an extra uncertainty component in $\sigma(\Delta_{\textrm{eff}})$ less than one percent.

\appendix

\section{Ramsey measurement}\label{appendix_ramsey_measurement}

\begin{figure}[b]
\centering
\includegraphics[width=9cm]{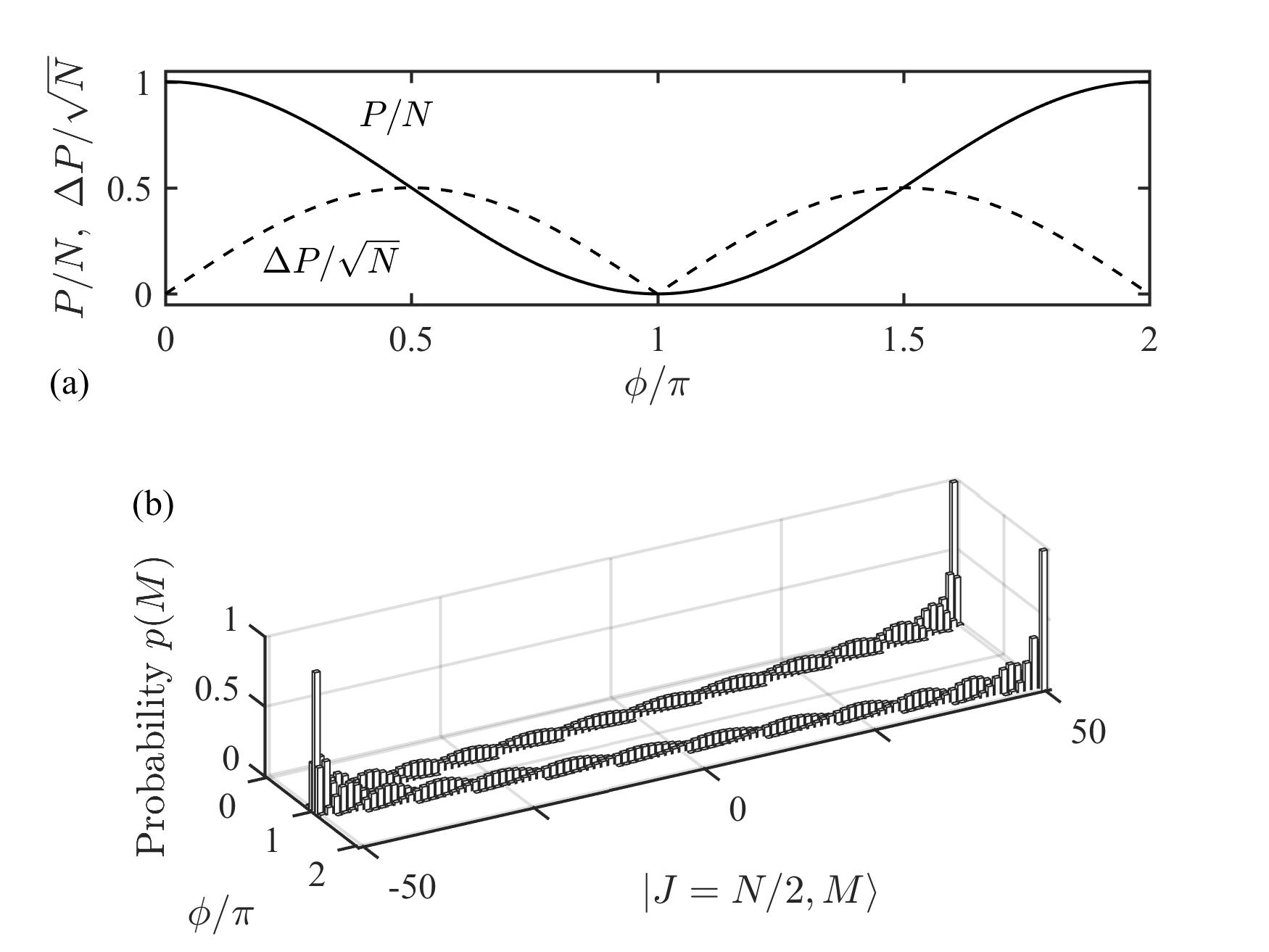}
\caption{Ramsey measurement. (a) Population $P$ in $\ket{\uparrow}$ and standard deviation $\Delta P$ as a function of the phase difference $\phi$ between two light pulses. (b) Distribution $p(M)$ of the system in $\ket{J=\frac{N}{2},M}$. For all curves, the spin number is $N=100$ and the system is initially prepared in $\ket{J=\frac{N}{2},M=-\frac{N}{2}}$.}
\label{fig8}
\end{figure}

We perform the Ramsey measurement on an ensemble of $N$ spins. In the Heisenberg picture, one has the projection operator
\begin{eqnarray}
\nonumber\hat{P}(\phi)&=&N+e^{i\pi\hat{J}_{x}/2}e^{-i\phi\hat{J}_{z}/2}e^{i\pi\hat{J}_{x}/2}\hat{J}_{z}\\
\nonumber&&\times e^{-i\pi\hat{J}_{x}/2}e^{i\phi\hat{J}_{z}/2}e^{-i\pi\hat{J}_{x}/2}\\
&=&N-(\hat{J}_{z}/2)\cos\phi-(\hat{J}_{x}/2)\sin\phi.
\end{eqnarray}
Here, $\phi$ is the phase difference between two light $\frac{\pi}{2}$-pulses. The expectation value of $\hat{P}(\phi)$ is given by $P(\phi)=\langle\hat{P}(\phi)\rangle=\bra{\psi}\hat{P}(\phi)\ket{\psi}$ with the initial state $\psi$ of the system. The standard deviation $\Delta P=\sqrt{\langle\hat{P}^{2}\rangle-\langle\hat{P}\rangle^{2}}$ of the projection measurement takes the form
\begin{eqnarray}
\nonumber(\Delta P)^{2}&=&2(\langle\hat{J}_{z}\hat{J}_{x}\rangle+\langle\hat{J}_{x}\hat{J}_{z}\rangle-2\langle\hat{J}_{z}\rangle\langle\hat{J}_{x}\rangle)\sin2\phi\\
&&+4(\Delta J_{z})^{2}\cos^{2}\phi+4(\Delta J_{x})^{2}\sin^{2}\phi,
\end{eqnarray}
with $\Delta J_{u=x,z}=\sqrt{\langle\hat{J}^{2}_{u}\rangle-\langle\hat{J}_{u}\rangle^{2}}$.

Figure~\ref{fig8}a illustrates the dependence of $P$ and $\Delta P$ on the phase difference $\phi$, where the system is initialized in $\ket{J=\frac{N}{2},M=-\frac{N}{2}}$. The standard deviation $\Delta P/\sqrt{N}$ reaches zero when $P/N=0$ (1), i.e., all spins are in $\ket{\downarrow}$ ($\ket{\uparrow}$) and is maximized at $P/N=1/2$. One may also compute the distribution $p(M)=|\langle J+N/2,M|\psi(\phi)\rangle|^{2}$ of spins in different states. Here $\psi(\phi)$ is the wavefunction of the system after the second light $\frac{\pi}{2}$-pulse is launched. We have $P=\sum_{M}(M+\frac{N}{2})p(M)$ and $(\Delta P)^{2}=\sum_{M}(M+\frac{N}{2})^{2}p(M)-P^{2}$. As depicted in Fig.~\ref{fig8}b, the distribution $p(M)$ spreads rapidly when $\psi(\phi)$ is away from $\ket{J=\frac{N}{2},M=\pm\frac{N}{2}}$.

According to $p(M)$ shown in Fig.~\ref{fig8}b, one may numerically simulate the spin excitation through the Monte Carlo method. For a certain $\phi$, we have to select a specific $(M+\frac{N}{2})$ with the corresponding probability $p(M)$ as a Ramsey measurement result. To this end, we compute the maximum $p_{\textrm{max}}$ in the array $\{p(-J),p(-J+1),\cdots,p(J-1),p(J)\}$, shuffle the elements in the array, and draw a uniform random number $x$ in $[0,1]$. Then, we compare $x$ to the first element $p(\tilde{M})$ in the shuffled array. If $x<\frac{p(\tilde{M})}{p_{\textrm{max}}}$, the Ramsey measurement result is $(\tilde{M}+\frac{N}{2})$. Otherwise, the procedure is repeated. One may obtain an ensemble of measurement results, whose mean value is $P$ and standard deviation is $\Delta P$.

\section{Enhanced sensitivity for detecting phase fluctuations}

\begin{figure}
\centering
\includegraphics[width=9cm]{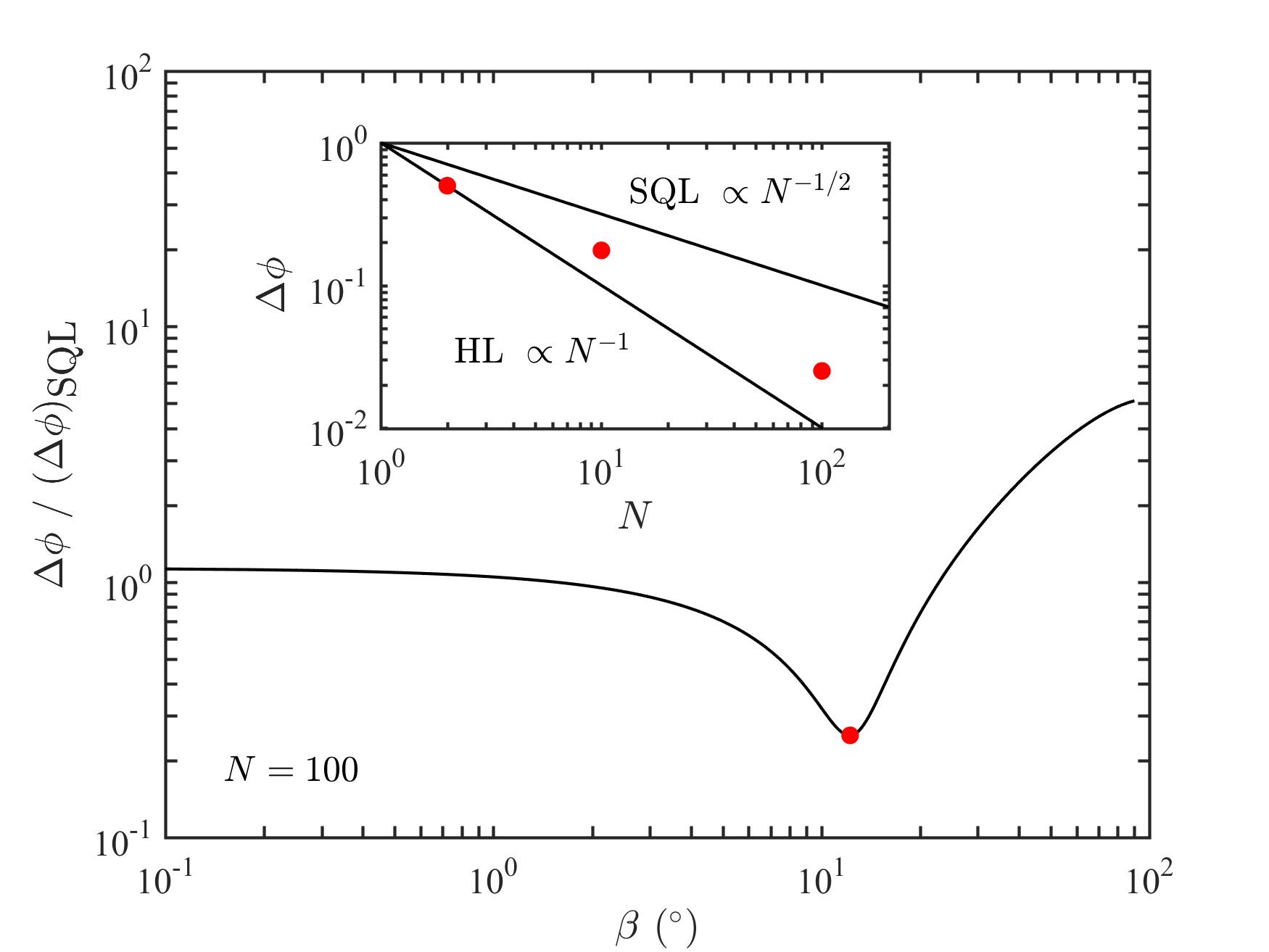}
\caption{Phase sensitivity $\Delta\phi$. The spin system is prepared in the entangled state $\psi_{\textrm{SSS}}(\alpha,\beta)$ with $N=100$ and $\alpha/\pi=1.016$. Varying $\beta$ changes $\Delta\phi$ and the minimum of $\Delta\phi$ (symbol $\bullet$) reaches $0.25(\Delta\phi)_{\textrm{SQL}}$. Inset: minimum of $\Delta\phi$ vs. $N$ (symbols $\bullet$).}
\label{fig9}
\end{figure}

The standard deviation $\Delta P$ of the projection measurement leads to the uncertainty $\Delta\phi$ of the phase detection. Usually, the phase difference $\phi$ is set to be $\frac{\pi}{2}$, which gives
\begin{eqnarray}
\nonumber\Delta\phi&=&\frac{\Delta P(\phi)}{\partial P(\phi)/\partial\phi}\\
&=&\frac{\Delta J_{x}}{|\langle\hat{J}_{z}\rangle|}.
\end{eqnarray}
When the spin system is initially prepared in $\ket{J=\frac{N}{2},M=-\frac{N}{2}}$, we have $\Delta J_{x}=\sqrt{N}/2$ and $|\langle\hat{J}_{z}\rangle|=N/2$ and arrive at the standard quantum limit
\begin{equation}\label{eq_sql}
(\Delta\phi)_{\textrm{SQL}}=1/\sqrt{N}.
\end{equation}
Equation~(\ref{eq_sql}) is also known as the quantum projection noise limit or the shot-noise limit.

Indeed, the standard quantum limit is not fundamental and can be exceeded using the entangled spins. Figure~\ref{fig9} shows an example of the uncertainty $\Delta\phi$ when the spin system is in
\begin{equation}\label{appendix_sss_eq}
\psi_{\textrm{SSS}}(\alpha,\beta)=e^{-i(\beta-\pi/2)\hat{J}_{y}}e^{-i\alpha\hat{J}_{z}^{2}}\psi_{\textrm{CSS}}(\theta=\pi/2,\textbf{n}=\textbf{e}_{x}),
\end{equation}
with $\psi_{\textrm{CSS}}(\theta,\textbf{n})=e^{-i\theta\textbf{n}\cdot\hat{\textbf{J}}}\ket{J=\frac{N}{2},M=-\frac{N}{2}}$ before the second light $\frac{\pi}{2}$-pulse is launched in the Ramsey measurement. It is seen that $\Delta\phi$ can be smaller than $(\Delta\phi)_{\textrm{SQL}}$ for some values of $\alpha$ and $\beta$. This is usually referred to as the spin squeezing~\cite{PhysRep:Ma2011} and has been employed in quantum metrology to improve the frequency stability of optical clocks~\cite{Nature:Pedrozo-Penafiel2020}.

The fundamental limit on the phase detection is imposed by the Heisenberg uncertainty relation. The commutation relation $[\hat{J}_{x},\hat{J}_{y}]=i\hat{J}_{z}$ gives the inequality $\Delta J_{x}\Delta J_{y}\geq|\langle\hat{J}_{z}\rangle|/2$. Since the maximum of $\Delta J_{y}$ is $J=N/2$, one obtains the Heisenberg limit on the phase detection
\begin{equation}
(\Delta\phi)_{\textrm{HL}}=1/N.
\end{equation}
For small spin systems, the entanglement operation [i.e., Eq.~(\ref{appendix_sss_eq})] may suppress $\Delta\phi$ to a value close to $(\Delta\phi)_{\textrm{HL}}$. As the system's size (i.e., the spin number $N$) is increased, the minimum of $\Delta\phi$ becomes far above $(\Delta\phi)_{\textrm{HL}}$ (see the inset in Fig.~\ref{fig9}).

\begin{acknowledgments}
D.Y. acknowledges the funding provided by National Time Service Center, China (E239SC11).
\end{acknowledgments}





\begin{thebibliography}{99}

\bibitem{PRL:Chou2010} C. W. Chou, D. B. Hume, J. C. J. Koelemeij, D. J. Wineland, and T. Rosenband, Frequency comparison of two high-accuracy ${\mathrm{Al}}^{+}$ optical clocks. Phys. Rev. Lett. {\bf 104}, 070802 (2010).

\bibitem{Science:Hinkley2013} N. Hinkley, J. A. Sherman, N. B. Phillips, M. Schioppo, N. D. Lemke, K. Beloy, M. Pizzocaro, C. W. Oates, and A. D. Ludlow, An Atomic Clock with $10^{-18}$ instability. Science {\bf 341}, 1215 (2013).

\bibitem{Nature:Bloom2014} B. J. Bloom, T. L. Nicholson, J. R. Williams, S. L. Campbell, M. Bishof, X. Zhang, W. Zhang, S. L. Bromley, and J. Ye, An optical lattice clock with accuracy and stability at the $10^{-18}$ level. Nature {\bf 506}, 71 (2014).

\bibitem{NatPhoton:Ushijima2015} I. Ushijima, M. Takamoto, M. Das, T. Ohkubo, and H. Katori, Cryogenic optical lattice clocks. Nat. Photon. {\bf 9}, 185 (2015).

\bibitem{PRL:Huntemann2016} N. Huntemann, C. Sanner, B. Lipphardt, Chr. Tamm, and E. Peik, Single-ion atomic clock with $3\ifmmode\times\else\texttimes\fi{}{10}^{\ensuremath{-}18}$ systematic uncertainty. Phys. Rev. Lett. {\bf 116}, 063001 (2016).

\bibitem{Metrologia:Bothwell2019} T. Bothwell, D. Kedar, E. Oelker, J. M. Robinson, S. L. Bromley, W. L. Tew, J. Ye, and C. J. Kennedy, JILA SrI optical lattice clock with uncertainty of $2.0\times 10^{-18}$. Metrologia {\bf 56}, 065004 (2019).

\bibitem{PRL:Huntemann2014} N. Huntemann, B. Lipphardt, Chr. Tamm, V. Gerginov, S. Weyers, and E. Peik, Improved limit on a temporal variation of ${m}_{p}/{m}_{e}$ from comparisons of ${\mathrm{Yb}}^{+}$ and Cs atomic clocks. Phys. Rev. Lett. {\bf 113}, 210802 (2014).

\bibitem{PRL:Godun2014} R. M. Godun, P. B. R. Nisbet-Jones, J. M. Jones, S. A. King, L. A. M. Johnson, H. S. Margolis, K. Szymaniec, S. N. Lea, K. Bongs, and P. Gill, Frequency ratio of two optical clock transitions in $^{171}{\mathrm{Yb}}^{+}$ and constraints on the time variation of fundamental constants. Phys. Rev. Lett. {\bf 113}, 210801 (2014).

\bibitem{NatPhys:Komor2014} P. K\' {o}m\' {o}r, E. M. Kessler, M. Bishof, L. Jiang, A. S. S\o{}rensen, J. Ye, and M. D. Lukin, A quantum network of clocks. Nat. Phys. {\bf 10}, 582 (2014).

\bibitem{NatPhoton:Takano2016} T. Takano, M. Takamoto, I. Ushijima, N. Ohmae, T. Akatsuka, A. Yamaguchi, Y. Kuroishi, H. Munekane, B. Miyahara, and H. Katori, Geopotential measurements with synchronously linked optical lattice clocks. Nat. Photon. {\bf 10}, 662 (2016).

\bibitem{NatCommun:Lisdat2016} C. Lisdat, et al. A clock network for geodesy and fundamental science. Nat. Commun. {\bf 7}, 12443 (2016).

\bibitem{PRL:Hutson2019} R. B. Hutson, A. Goban, G. E. Marti, L. Sonderhouse, C. Sanner, and J. Ye, Engineering quantum states of matter for atomic clocks in shallow optical lattices. Phys. Rev. Lett. {\bf 123}, 123401 (2019).

\bibitem{Science:Aeppli2022} A. Aeppli, A. Chu, T. Bothwell, C. J. Kennedy, D. Kedar, P. He, A. M. Rey, J. Ye, Hamiltonian engineering of spin-orbit–coupled fermions in a Wannier-Stark optical lattice clock. Science {\bf 8}, eadc9242 (2022).

\bibitem{Science:Swallows2011} M. D. Swallows, M. N. Bishof, Y. Lin, S. Blatt, M. J. Martin, A. M. Rey, and J. Ye, Suppression of collisional shifts in a strongly interacting lattice clock. Science {\bf 331}, 1043 (2011).

\bibitem{NatPhoton:Takamoto2020} M. Takamoto, I. Ushijima, N. Ohmae, T. Yahagi, K. Kokado, H. Shinkai, and H. Katori, Test of general relativity by a pair of transportable optical lattice clocks. Nat. Photon. {\bf 14}, 411 (2020).

\bibitem{IEEE:Santarelli1998} G. Santarelli, C. Audoin, A. Makdissi, P. Laurent, G. J. Dick, and A. Clairon, Frequency stability degradation of an oscillator slaved to a periodically interrogated atomic resonator. IEEE Trans. Ultrason. Ferroelectr. Freq. Control {\bf 45}, 887 (1998).

\bibitem{NatPhoton:Kessler2012} T. Kessler, C. Hagemann, C. Grebing, T. Legero, U. Sterr, F. Riehle, M. J. Martin, L. Chen, and J. Ye, A sub-40-mHz-linewidth laser based on a silicon single-crystal optical cavity. Nat. Photon. 6, 687 (2012).

\bibitem{Optica:Robinson2019} J. M. Robinson, E. Oelker, W. R. Milner, W. Zhang, T. Legero, D. G. Matei, F. Riehle, U. Sterr, and J. Ye, Crystalline optical cavity at 4 K with thermal-noise-limited instability and ultralow drift. Optica {\bf 6}, 240 (2019).

\bibitem{PRL:Matei2017} D. G. Matei, T. Legero, S. H\"afner, C. Grebing, R. Weyrich, W. Zhang, L. Sonderhouse, J. M. Robinson, J. Ye, F. Riehle, and U. Sterr, $1.5\text{ }\text{ }\ensuremath{\mu}\mathrm{m}$ lasers with sub-10 mHz linewidth. Phys. Rev. Lett. {\bf 118}, 263202 (2017).

\bibitem{NatPhoton:Takamoto2011} M. Takamoto, T. Takano, and H. Katori, Frequency comparison of optical lattice clocks beyond the Dick limit. Nat. Photon. {\bf 5}, 288 (2011).

\bibitem{NatPhoton:Schioppo2017} M. Schioppo, R. C. Brown, W. F. McGrew, N. Hinkley, R. J. Fasano, K. Beloy, T. H. Yoon, G. Milani, D. Nicolodi, J. A. Sherman, N. B. Phillips, C. W. Oates, and A. D. Ludlow, Ultrastable optical clock with two cold-atom ensembles. Nat. Photon. {\bf 11}, 48 (2017).

\bibitem{NatPhoton:Oelker2020} E. Oelker, R. B. Hutson, C. J. Kennedy, L. Sonderhouse, T. Bothwell, A. Goban, D. Kedar, C. Sanner, J. M. Robinson, G. E. Marti, D. G. Matei, T. Legero, M. Giunta, R. Holzwarth, F. Riehle, U. Sterr, and J. Ye, Demonstration of $4.8\times10^{-17}$ stability at 1 s for two independent optical clocks. Nat. Photon. {\bf 13}, 714 (2019).

\bibitem{Nature:Zheng2022} X. Zheng, J. Dolde, V. Lochab, B. N. Merriman, H. Li, and S. Kolkowitz, Differential clock comparisons with a multiplexed optical lattice clock. Nature {\bf 602}, 425 (2022).

\bibitem{Nature:Bothwell2022} T. Bothwell, C. J. Kennedy, A. Aeppli, D. Kedar, J. M. Robinson, E. Oelker, A. Staron, and J. Ye, Resolving the gravitational redshift across a millimetre-scale atomic sample. Nature {\bf 602}, 420 (2022).

\bibitem{NatCommun:Schulte2020} M. Schulte, C. Lisdat, P. O. Schmidt, U. Sterr, and K. Hammerer, Prospects and challenges for squeezing-enhanced optical atomic clocks. Nat. Commun. {\bf 11}, 5955 (2020).

\bibitem{NatPhys:Schine2022} N. Schine, A. W. Young, W. J. Eckner, M. J. Martin, and A. M. Kaufman, Long-lived Bell states in an array of optical clock qubits. Nat. Phys. {\bf 18}, 1067 (2022).

\bibitem{Nature:Nichol2022} B. C. Nichol, R. Srinivas, D. P. Nadlinger, P. Drmota, D. Main, G. Araneda, C. J. Ballance, and D. M. Lucas, An elementary quantum network of entangled optical atomic clocks. Nature {\bf 609}, 689 (2022).

\bibitem{Nature:Pedrozo-Penafiel2020} E. Pedrozo-Pe\~nafiel, S. Colombo, C. Shu, A. F. Adiyatullin, Z. Li, E. Mendez, B. Braverman, A. Kawasaki, D. Akamatsu, Y. Xiao, and V. Vuleti\ifmmode \acute{c}\else \'{c}\fi{}, Entanglement on an optical atomic-clock transition. Nature {\bf 588}, 414 (2020).

\bibitem{Science:Wciso2018} P. Wcis\l{}o, et al. New bounds on dark matter coupling from a global network of optical atomic clocks. Sci. Adv. {\bf 4}, eaau4869 (2018).

\bibitem{PRD:Kolkowitz2016} S. Kolkowitz, I. Pikovski, N. Langellier, M. D. Lukin, R. L. Walsworth, and J. Ye, Gravitational wave detection with optical lattice atomic clocks. Phys. Rev. D {\bf 94}, 124043 (2016). 

\bibitem{SciRep:Gao2018} Q. Gao, M. Zhou, C. Han, S. Li, S. Zhang, Y. Yao, B. Li, H. Qiao, D. Ai, G. Lou, M. Zhang, Y. Jiang, Z. Bi, L. Ma, and X. Xu, Systematic evaluation of a 171Yb optical clock by synchronous comparison between two lattice systems. Sci. Rep. {\bf 8}, 8022 (2018).

\bibitem{PRL:Marti2018} G. E. Marti, R. B. Hutson, A. Goban, S. L. Campbell, N. Poli, and J. Ye, Imaging optical frequencies with $100\text{ }\text{ }\ensuremath{\mu}\mathrm{Hz}$ precision and $1.1\text{ }\text{ }\ensuremath{\mu}\mathrm{m}$ resolution. Phys. Rev. Lett. {\bf 120}, 103201 (2018).

\bibitem{PRA:Porsev2004} S. G. Porsev, A. Derevianko, and E. N. Fortson, Possibility of an optical clock using the $6{}^{1}{S}_{0}\ensuremath{\rightarrow}6{}^{3}{P}_{0}^{o}$ transition in ${}^{171,173}\mathrm{Yb}$ atoms held in an optical lattice. Phys. Rev. A {\bf 69}, 021403 (2004).

\bibitem{PRA:Itano1993} W. M. Itano, J. C. Bergquist, J. J. Bollinger, J. M. Gilligan, D. J. Heinzen, F. L. Moore, M. G. Raizen, and D. J. Wineland, Quantum projection noise: Population fluctuations in two-level systems. Phys. Rev. A {\bf 47}, 3554 (1993).

\bibitem{PhysRep:Ma2011} J. Ma, X. Wang, C. P. Sun, and F. Nori, Quantum spin squeezing. Phys. Rep. {\bf 509}, 89 (2011).

\bibitem{PRA:Santarelli1999} G. Santarelli, Ph. Laurent, P. Lemonde, A. Clairon, A. G. Mann, S. Chang, A. N. Luiten, and C. Salomon, Quantum projection noise in an atomic fountain: A high stability cesium frequency standard. Phys. Rev. Lett. {\bf 82}, 4619 (1999).

\bibitem{PRA:Wineland1992} D. J. Wineland, J. J. Bollinger, W. M. Itano, F. L. Moore, and D. J. Heinzen, Spin squeezing and reduced quantum noise in spectroscopy. Phys. Rev. A {\bf 46}, R6797 (1992).

\bibitem{PRA:Wineland1994} D. J. Wineland, J. J. Bollinger, W. M. Itano, and D. J. Heinzen, Squeezed atomic states and projection noise in spectroscopy. Phys. Rev. A {\bf 50}, 67 (1994).

\bibitem{PRA:Kitagawa1993} M. Kitagawa and M. Ueda, Squeezed spin states. Phys. Rev. A {\bf 47}, 5138 (1993).

\bibitem{PRA:SchleierSmith2010} M. H. Schleier-Smith, I. D. Leroux, and Vuleti\ifmmode \acute{c}\else \'{c}\fi{}, Squeezing the collective spin of a dilute atomic ensemble by cavity feedback. Phys. Rev. A {\bf 81}, 021804 (2010).

\bibitem{PRL:Perlin2020} M. A. Perlin, C. Qu, and A. M. Rey, Spin squeezing with short-range spin-exchange interactions. Phys. Rev. Lett. {\bf 125}, 223401 (2020).

\bibitem{Science:Lange2018} K. Lange, J. Peise, B. L\"ucke, I. Kruse, G. Vitagliano, I. Apellaniz, M. Kleinmann, G. T\'oth, and C. Klempt, Entanglement between two spatially separated atomic modes. Science {\bf 360}, 416 (2018).

\bibitem{Nature:Bao2020} H. Bao, J. Duan, S. Jin, X. Lu, P. Li, W. Qu, M. Wang, I. Novikova, E. E. Mikhailov, K.-F. Zhao, K. M\o{}lmer, H. Shen, and Y. Xiao, Spin squeezing of $10^{11}$ atoms by prediction and retrodiction measurements. Nature {\bf 581}, 159 (2020).

\bibitem{PRX:Madjarov2019} I. S. Madjarov, A. Cooper, A. L. Shaw, J. P. Covey, V. Schkolnik, T. H. Yoon, J. R. Williams, and M. Endres, An atomic-array optical clock with single-atom readout. Phys. Rev. X {\bf 9}, 041052 (2019).

\bibitem{Science:Norcia2019} M. A. Norcia, A. W. Young, W. J. Eckner, E. Oelker, J. Ye, and A. M. Kaufman, Seconds-scale coherence on an optical clock transition in a tweezer array. Science {\bf 366}, 93 (2019).

\bibitem{PRL:Covey2019} J. P. Covey, I. S. Madjarov, A. Cooper, and M. Endres, 2000-times repeated imaging of strontium atoms in clock-magic tweezer arrays. Phys. Rev. Lett. {\bf 122}, 173201 (2019).












\end{thebibliography}
\end{document}